\documentclass[letterpaper,10pt,conference]{ieeeconf}

\usepackage{graphicx}
\usepackage{amsmath}
\usepackage{amssymb}
\usepackage{url}
\usepackage{multirow}
\usepackage{booktabs}
\usepackage{subcaption}

\usepackage{pifont}
\usepackage{bm}
\let\labelindent\relax \usepackage[inline]{enumitem}
\usepackage{makecell}
\graphicspath{{./figs/}}

\IEEEoverridecommandlockouts

 \title{A Bottom-up method Towards the Automatic
  and Objective Monitoring of Smoking Behavior In-the-wild using
  Wrist-mounted Inertial Sensors}

\author{
  Athanasios Kirmizis,
  Konstantinos Kyritsis and Anastasios Delopoulos
  \thanks{All authors are with the Multimedia Understanding Group, Information Processing Laboratory, Aristotle University of Thessaloniki, Greece.}%
  \thanks{$\copyright$ 2021 IEEE.  Personal use of this material is permitted. Permission from IEEE must be obtained for all other uses, in any current or future media, including reprinting/republishing this material for advertising or promotional purposes, creating new collective works, for resale or redistribution to servers or lists, or reuse of any copyrighted component of this work in other works.}
}

\makeatletter
\let\NAT@parse\undefined
\newcommand*{\rom}[1]{\expandafter\@slowromancap\romannumeral #1@}
\makeatother

\usepackage{cite}
\usepackage{hyperref}

\hypersetup{
bookmarksnumbered,
pdfstartview={FitH},
citecolor={blue},
linkcolor={blue},
urlcolor={black},
pdfpagemode={UseOutlines}
}

\begin{document}

\maketitle
\thispagestyle{empty}
\pagestyle{empty}


\begin{abstract}

The consumption of tobacco has reached global epidemic proportions and
is characterized as the leading cause of death and illness. 
Among the different ways of consuming tobacco
(e.g., smokeless, cigars), smoking cigarettes is the most
widespread. In this paper, we present a two-step, bottom-up algorithm
towards the automatic and objective monitoring of cigarette-based,
smoking behavior during the day, using the 3D acceleration and
orientation velocity measurements from a commercial smartwatch. In the
first step, our algorithm performs the detection of individual smoking
gestures (i.e., puffs) using an artificial neural network with both
convolutional and recurrent layers. In the second step, we make use of
the detected puff density to achieve the temporal localization of
smoking sessions that occur throughout the day. In the experimental
section we provide extended evaluation regarding each step of the
proposed algorithm, using our publicly-available, realistic Smoking
Event Detection (SED) and Free-living Smoking Event Detection (SED-FL)
datasets recorded under semi-controlled and free-living conditions,
respectively. In particular, leave-one-subject-out (LOSO) experiments
reveal an F1-score of 0.863 for the detection of puffs and an
F1-score/Jaccard index equal to 0.878/0.604 towards the temporal
localization of smoking sessions during the day. Finally, to gain
further insight, we also compare the puff detection part of our
algorithm with a similar approach found in the recent literature.
  
\end{abstract}


\section{Introduction}
\label{sec:intro}

According to the World Health Organization (WHO), smoking is the
leading public health problem worldwide, resulting in millions of
preventable deaths each year and is responsible for a number of
serious chronic diseases (e.g., hypertension, atherosclerosis, cancer)
\cite{world2017report}. 

Globally, male and female smokers have their life expectancy reduced
by $13.2$ and $14.5$ years, respectively
\cite{centers2002annual}. Moreover, at least half of all smokers
worldwide die prematurely from smoking \cite{world2017report}. 
It is important to emphasize that smoking is
not only harmful to smokers themselves, but it is also a major risk
factor for passive smokers \cite{otsuka2001acute}.

The modernization of societies has ignited a recent trend that
promotes a lifestyle in which smoking is considered an outdated
habit. More and more people are taking up a sport, or are beginning to
follow a healthy eating regime \cite{global2018}. An important role
for the engagement of people with these healthy habits plays the
technology that is constantly evolving and gets integrated into
everyday life. The rapid growth of portable and wearable devices has
brought with it a great increase in applications that help people
develop and maintain a healthy lifestyle. From tracking meals and
calories to measuring physical activity or sleep, the applications now
available to users are multiple, easy to use, and unobtrusive
\cite{west2012there}. However, the objective monitoring of smoking
behavior is still an open research problem. Research
shows \cite{lancaster2005self} that tailored feedback to the smoker
can greatly facilitate the reduction or even permanent cessation of
smoking.

Several works exist in the literature that approach the problem of
smoking behavior monitoring using body-worn sensors
\cite{imtiaz2019wearable}. The work presented in
\cite{saleheen2015puffmarker} suggests a method that combines the data
from a wrist-mounted inertial measurement unit (IMU) sensor and a
chest-worn respiratory inductive plethysmography (RIP) sensor towards
the detection of smoking gestures. Evaluation using data from $6$
daily smokers reveals a recall of $0.97$. It should be mentioned,
however, that devices as bulky as the RIP sensor are too obtrusive for
the user to properly simulate the normal smoking behavior.
The work of M. Shoaib \textit{et al.} \cite{shoaib2016hierarchical}
proposes a two-step algorithm towards the detection of smoking
events. In the first stage, the data are crudely classified, while at
the second stage a rule-based correction of the first-stage
classification is applied. The classifiers tested by the authors are
random forest (RF), decision tree (DT) and support vector machine
(SVM). For each classifier, a total of $36$ features are extracted
from the $3$D accelerometer and gyroscope measurements. According to
the authors, the second step of their algorithm corrects up to $50$\%
of the misclassified samples. Evaluation is performed using their
dataset of $11$ participants with a total duration of $45$ hours,
where the authors achieved an F$1$-score of $0.83$-$0.94$.

\begin{figure*}[h]
  \centering
  \includegraphics[width=\textwidth]{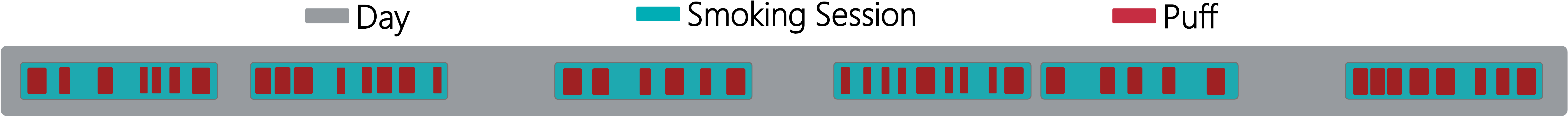}
  \caption{\small{The proposed smoking behavior model. In this example, a
    subject performed six smoking sessions (blue) during the course of
    the day (grey), with each session containing a number of puffs
    (dark red).}}
\label{fig:model}
\end{figure*}

In our work, we propose the use of a smoking behavior model that is
based on two fundamental components: a) the \textit{puff} (also referred to as smoking gesture in the literature), defined as
the series of hand movements that bring an active cigarette to the
mouth with the purpose of smoking it and then back to rest, and b) the
\textit{smoking session}, defined as the act of consuming a
cigarette. In particular, we model smoking behavior as a series of
smoking sessions that occur during the day. Subsequently, each smoking
session is modeled as a series of puffs. Figure \ref{fig:model}
illustrates the adopted smoking behavior model. Furthermore, we
suggest a two-step, bottom-up method towards the objective and
automatic monitoring of smoking behavior using all-day, free-living
IMU recordings from an off-the-shelf smartwatch. In the first step, we
use an artificial neural network (ANN) with convolutional and
recurrent layers to detect puffs during a smoking session. In the
second step, we use the distribution of the detected puffs to localize
the smoking sessions throughout the day.


\section{Detection of puffs}
\label{sec:method}


\subsection{Data pre-processing}
\label{sec:preproc}

Let $\bm{s}(t) = [\bm{a}_x(t), \bm{a}_y(t), \bm{a}_z(t),\bm{g}_x(t),
  \bm{g}_y(t), \bm{g}_z(t)]^T$ represent the vector that contains the
$3$D acceleration and orientation velocity measurements for a moment
$t$. Then, a complete recording of \(d_{tot}\) seconds can be
represented by the $M\times6$ signal \(\bm{R} =
[\bm{s}(1),\bm{s}(2),\ldots,\bm{s}(M)]^T\), where \(M = d_{tot} \cdot
f_s\) is the length of the recording in samples and \(f_s\) is the
sampling frequency in Hz.

Smoking cigarettes is a process that can be completed by using either
hand (right or left) or, in some cases, a combination of both. In
order to achieve uniformity among data from different participants, we
consider the right hand as the reference and transform all left-handed
smoking sessions using the \textit{hand mirroring} process proposed
by Kyritsis \textit{et al.} \cite{kyritsis2020data}. Particularly, all
recordings that are collected with the participant wearing the
smartwatch on the left wrist $\bm{R}_{l}$, are transformed into
$\bm{R}_{r}$ by changing the direction of the first, fifth and sixth
channels (i.e., $\bm{a}_{x}$, $\bm{g}_{y}$ and $\bm{g}_{z}$) of
$\bm{R}_{l}$.

Furthermore, accelerometer measurements also include the influence of
the Earth's gravitational field. To attenuate this undesirable effect,
a high-pass finite impulse response (FIR) filter is applied to each of
the acceleration streams (i.e., the first, second, and third channels
of $\bm{R}$), independently. Experimentally, we obtained satisfactory
results with a cut-off frequency of $1$ Hz and a filter length equal
to $512$ samples (which corresponds to \(512 /f_s\) seconds).


\subsection{Training the puff detection model}
\label{sec:model}
Given a recording $\bm{R}$ that corresponds to a smoking session, we
extract training examples using a sliding window. More
specifically, the sliding window has a length $w_l$ that corresponds
to $4.5$ seconds ($4.5 \, f_s$ samples) and a step $w_s$ that
corresponds to $0.5$ seconds ($0.5 \, f_s$ samples). We selected a
window length equal to $4.5$ seconds as it approximates the median
puff duration in the SED dataset (Table \ref{table:stats1}). Each
extracted window $\bm{W}_{i}$ has dimensions $(4.5 \, f_s) \times 6$.

\begin{table}[h]
\caption{\small{Information regarding the SED and SED-FL datasets. Statistics were calculated using the raw data.}}
\label{table:stats1}
\resizebox{0.48\textwidth}{!}{\begin{tabular}{l|c c |c c} 
\toprule
Dataset\ & \multicolumn{2}{c|}{\textbf{SED}} & \multicolumn{2}{c}{\textbf{SED-FL}} \\ \midrule
Session\ & {Smoking } & {Puffs} & {In-the-wild} & {Smoking } \\
\midrule
{Number of instances} & 20 & 276 & 10 & 39\\ 
{Mean (sec)} & 485.14 & 4.86 & 28202.14 & 525.33\\
{Std (sec)} & 197.32 & 1.47 & 13484.42 & 301.81\\
{Median (sec)} & 484.26 & 4.75 & 25919.27 & 462.80\\
{Total (sec)} & 9702.88 & 1341.18 & 282021.38 & 20487.84\\
{Total (hours)} & 2.69 & 0.37 & 78.33 & 5.69\\
\midrule
{Participants} & \multicolumn{2}{c|}{11} & \multicolumn{2}{c}{7} \\
\bottomrule
\end{tabular}}

\end{table}

\begin{figure}[h]
\centering
\includegraphics[width=\linewidth]{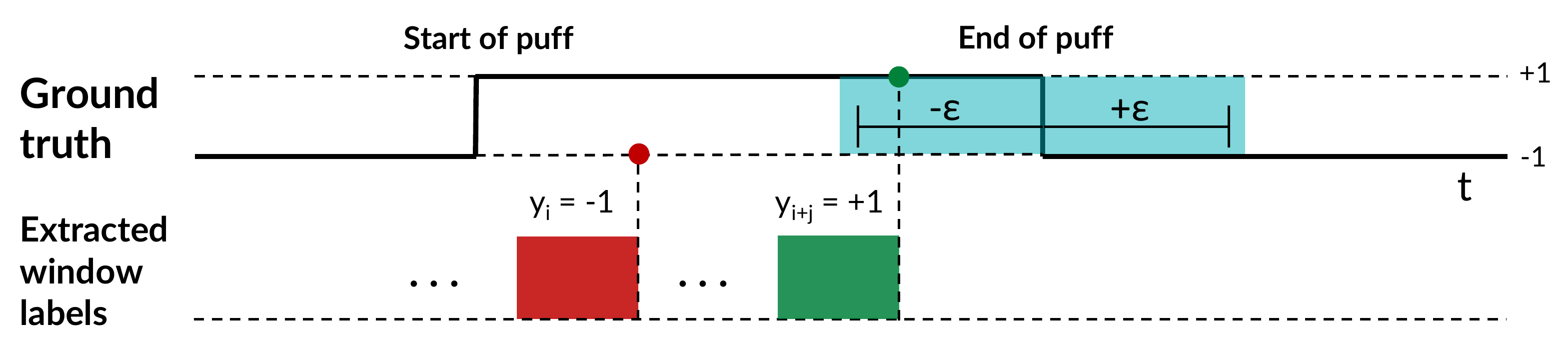} 
\caption{\small{The proposed window labeling method.}}
\label{fig:labeling}
\end{figure}

In order to train the network, each window $\bm{W}_{i}$ needs to be
associated with a label $y_{i}$ that would indicate if the window
corresponds to a puff or not ($y_{i} = \pm1$, respectively). We use
the following formula to perform the labeling process:

\begin{equation}
  y_i =
  \begin{cases}
       +1 & \text{if} \ t_j^{gt} - \epsilon \leq t_i^{\bm{W}} \leq t_j^{gt} + \epsilon \\
       -1 & \text{otherwise} 
  \end{cases}
\end{equation}
where $t_j^{gt}$ is the moment at which the $j$-th puff ends (hand has
returned to rest) according to ground truth (GT) and $t_i^{\bm{W}}$ is
the timestamp associated with the right end of the $i$-th extracted
window. Moreover, we select $\epsilon$ to be equal to $1.5$
seconds. Figure \ref{fig:labeling} showcases the window labeling
process.

The next step is to artificially augment the training set by
simulating different positions of the smartwatch, that may occur
\textit{involuntarily} while wearing it, with respect to the subject's
wrist. Specifically, we draw two numbers from a normal distribution with a mean and standard deviation
equal to $0$ and $10$, respectively. These two numbers represent the
\textit{angles} that the
smartwatch has rotated around the $x$ (parallel to the subject's arm)
and $z$ (perpendicular to the screen of the smartwatch) axes. The
transformation for each window $\bm{W}_{i}$ is selected to be one of
the following (with equal probability): a) rotation around $x$, b)
rotation around $z$, c) rotation around $x$ and then around $z$, or d)
rotation around $z$ and then around $x$. The motivation behind the
augmentation step was the significant increase in the performance
reported in \cite{kyritsis2020data}.

The proposed model is a tuned-down version of the renowned VGG
architecture \cite{simonyan2014very}. In particular, our network
includes a convolutional and a recurrent part. The convolutional part
contains three 1D convolutional layers, with each of the first two
followed by a max pooling layer with a decimation factor of $2$. The
convolutional layers have $32$, $64$ and $128$ filters, with a size of
$5$, $3$ and $3$, respectively. All convolutional layers use a unary
stride and the rectified linear unit (ReLU) as the non-linearity. The
recurrent part of the network consists of a single
long-short-term-memory (LSTM) layer with $128$ cells and the sigmoid
function as the activation of the recurrent steps. The output of the
LSTM is propagated to a fully connected layer with a single neuron and
the sigmoid activation function. In order to avoid overfitting, we
apply dropout to the inputs of the fully connected layer with a
probability of 50\%. The network minimizes the binary cross-entropy
loss with the RMSProp optimizer, and uses a learning rate of
$10^{-3}$, a batch size of $32$ and a number of $10$ epochs. In a
compact notation, the network can be written as
Conv($32\times5$)-Pool($2$)-Conv($64\times3$)-Pool($2$)-Conv($128\times3$)-LSTM($128$)-FC($1$),
where Conv($32\times 5$) represents a convolutional layer with $32$
filters and a filter size of $5$, Pool($2$) is pooling layer with a
decimation factor of $2$, LSTM($128$) is an LSTM layer with $128$
hidden cells and FC($1$) is a fully connected layer with a single
neuron.


\subsection{Puff detection}
By forwarding windows from a recording $\bm{R}$ to the trained
puff detection network (Section \ref{sec:model}), we obtain the predictions vector
$\bm{p}$ with length $N$. Essentially $\bm{p}_i$ is the probability
that the $i$-th window $\bm{W}_i$ is a puff and $N$ represents the total number
of extracted windows of length $w_l$ and step $w_s$.

\begin{figure}[h]
\centering
\includegraphics[width=\linewidth]{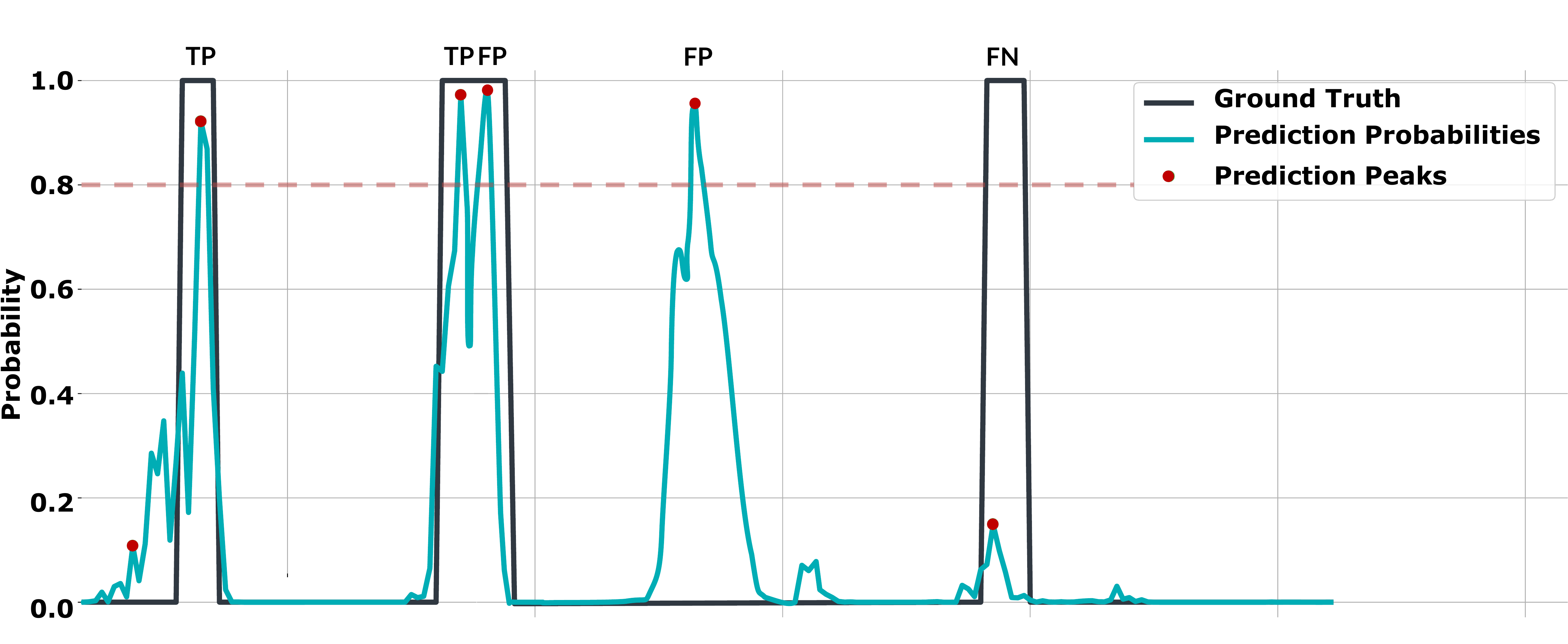} 
\caption{\small{Detection of puffs given the probability vector
    $\bm{p}$ (blue line). The ground truth puff durations (black
    line), local maxima peaks (red dots) and the $\lambda_p$ threshold
    (light-red dashed line) are also depicted. In the figure's
    example, the left-most and right-most peaks are rejected as they
    are below the threshold $\lambda_p$.}}
\label{fig:detection1}
\end{figure}

Puff detection is achieved by initially performing a local
maxima search in $\bm{p}$, with a minimum distance between successive
peaks equal to $10$ samples. The next step is to discard peaks that
are associated with a probability $\bm{p}_{i}$ that is lower than a
threshold $\lambda_p$ set to $0.8$. Both the minimum distance between peaks and $\lambda_p$ were selected by experimenting with a small part of the SED dataset. As a result, we obtain the set of detected
puffs, $\mathcal{F} = \{ f_{1}, \ldots, f_{K} \}$, where
$f_{i}$ is the timestamp of $i$-th detected puff and $K$ the total number of detected puffs. The process is illustrated in Figure
\ref{fig:detection1}.

\section{Temporal localization of smoking sessions}
\label{sec:method2}

The second step of the proposed algorithm aims at the temporal
localization of smoking sessions that occur during a day. In our early
experiments we observed that in all-day recordings the density of
puffs is increased during a smoking session and reduced everywhere
else. As a result, in the second step of our algorithm we take
advantage of this observation and attempt to group the detected puffs
into smoking session clusters using the density-based spatial
clustering of applications with noise (DBSCAN) \cite{ester1996density}
algorithm.

More specifically, let $\bm{R}'$ be an all-day, in-the-wild recording
with dimensions $M'\times6$, where $M' \gg M$. Next, we use the
trained puff detection model (Section \ref{sec:model}) to produce
the set of puff detection estimates $\mathcal{F'}$. Subsequently, we
apply clustering using DBSCAN on the set $\mathcal{F'}$ using a
minimum distance between clusters that corresponds to $250$ seconds (as this is the minimum distance between consecutive smoking sessions in the SED-FL dataset) and a
minimum number of points per cluster set to $4$.

\begin{figure}[h]
\centering \includegraphics[width=\linewidth]{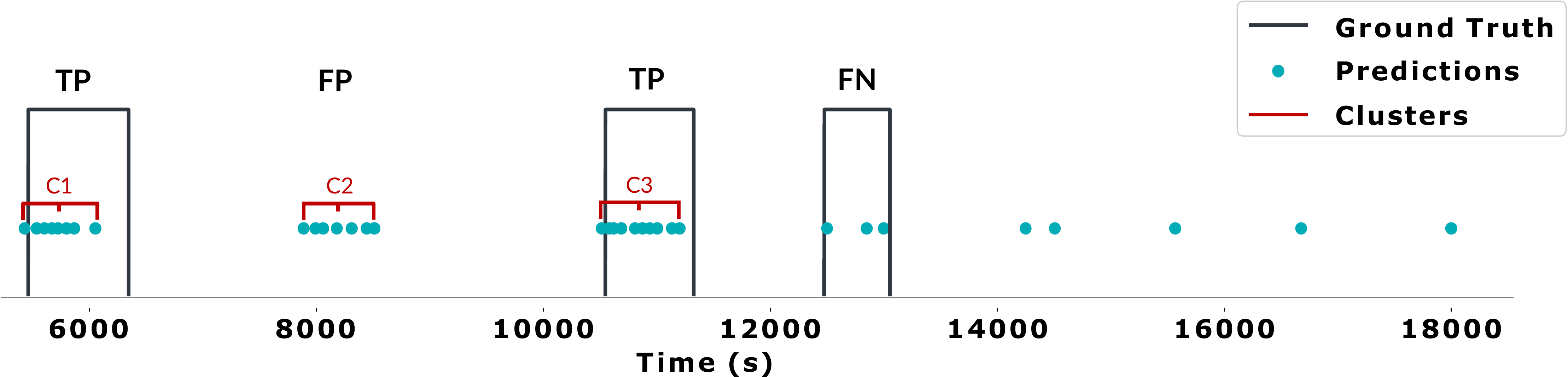}
\caption{\small{Figure depicting an example of how the smoking session
    clusters (red line) are formed using the set of detected puffs
    $\mathcal{F}'$ (blue dots). The ground truth smoking session
    durations (as annotated by the participants are also depicted
    (black line).}}
\label{fig:density}
\end{figure}

Each cluster that DBSCAN produces is then associated with the first
and last timestamps of the detected puffs that belong to that specific
cluster. This pair of timestamps corresponds to the start and end
moments of a smoking session. Formally, the final output of the
algorithm is the set $\mathcal{G} = \{ C_{1}, \ldots, C_{L} \} =
\{[t_{1}^{s},t_{1}^{e}], \ldots, [t_{L}^{s},t_{L}^{e}]\}$, where
$[t_{i}^{s},t_{i}^{e}]$ represents the start and end timestamps of the
$i$-th detected smoking session. An example depicting the temporal
localization of smoking sessions can be found in Figure
\ref{fig:density}.

\section{Experiments and evaluation}
\label{sec:experiments_results}


\subsection{Datasets}
\label{sec:dataset}

In order to fine-tune and evaluate our method we collected two
datasets. The SED dataset was captured in semi-controlled environments
(e.g., private residences or cafes) and contains a single smoking
session per recording. On the other hand, the SED-FL dataset was
captured under in-the-wild conditions and contains all-day recordings
that include smoking sessions and other daily activities (e.g.,
working, eating). Inertial data were collected using a Mobvoi TicWatch
E smartwatch at a sampling rate $f_s$ equal to $50$ Hz.

\begin{figure}[h]
\centering
\includegraphics[width=1\linewidth]{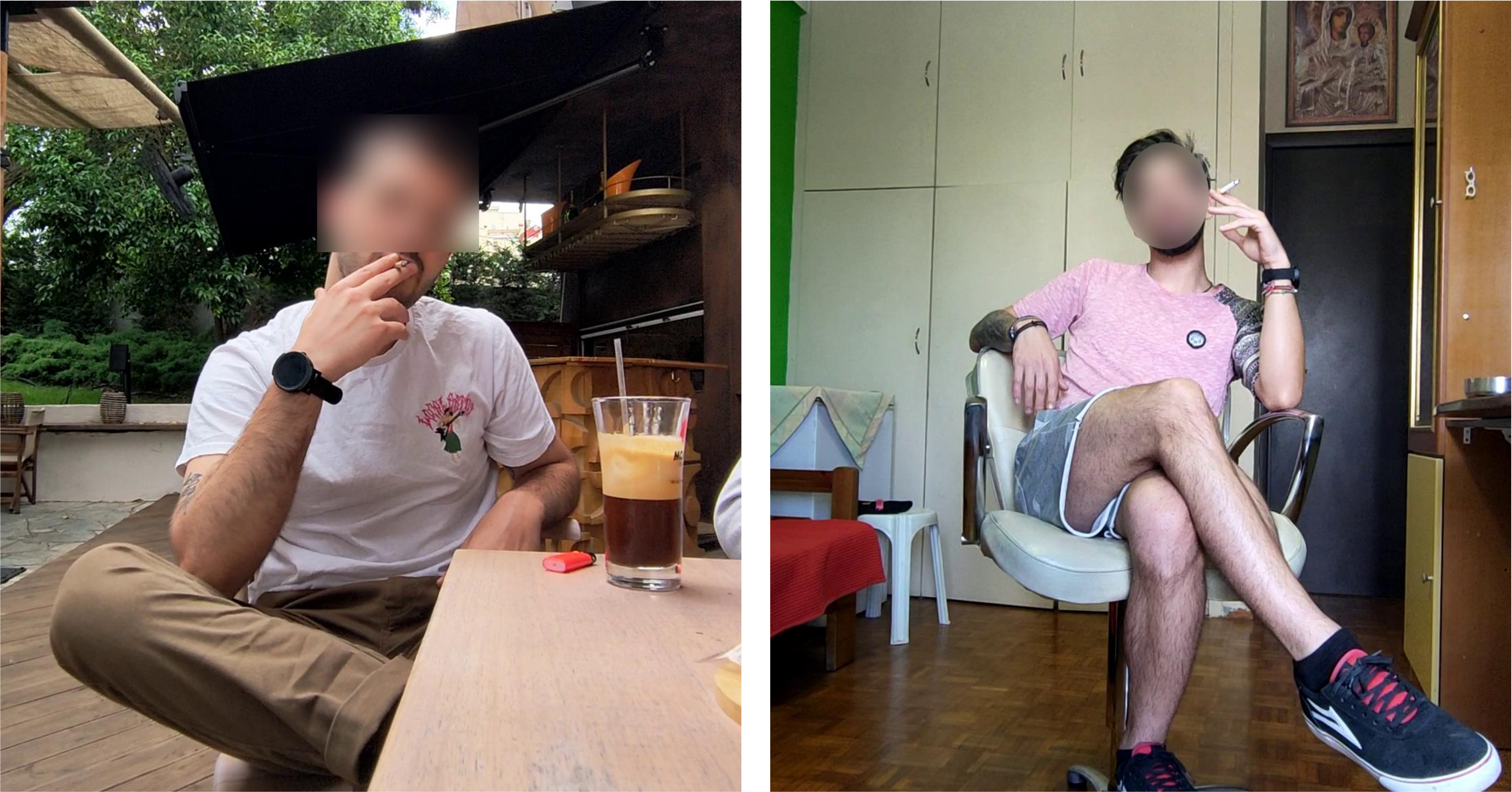} 
\caption{\small{Two participants from the SED dataset.}}
\label{fig:faces}
\end{figure}

The SED dataset consists of $11$ subjects performing $20$ smoking
sessions, with a total duration of $2.69$ hours. The SED-FL dataset
consists of $10$ all-day sessions from $7$ subjects, with a total
duration of $78.3$ hours (Table \ref{table:stats1}). Three of the
subjects participate in both datasets. It should be emphasized that we
asked from the subjects to smoke naturally; as a result, they were
free to engage in a discussion or perform additional activities (two
instances are depicted in Figure \ref{fig:faces}). All subjects were
already smokers and signed an informed consent prior to their
participation. In order to label the data in SED, we recorded the
smoking sessions using the camera from a typical smartphone. To
produce the GT for the all-day, in-the-wild sessions of SED-FL, a
smartwatch application was developed that enabled subjects to easily
note the start and end timestamps of their smoking sessions. It is
worth noting that both datasets deal with the consumption of tobacco
using \textit{cigarettes}; no electronic cigarettes (also known as
vaping devices), pipes or heated tobacco products were used. Both
datasets are publicly available at
\url{https://mug.ee.auth.gr/smoking-event-detection/}.

\subsection{Experiments}
\label{sec:experiments}

We conducted two series of experiments. In the first experiment
(EX-I), we evaluate the puff detection performance using the SED
dataset. Moreover, we compare the performance of the proposed puff
detection approach with the method proposed in
\cite{shoaib2016hierarchical}. For the second experiment (EX-II), we
evaluated the smoking session temporal localization performance using
the SED-FL dataset. Both EX-I and EX-II are performed in a
leave-one-subject-out (LOSO) fashion.


\subsection{Evaluation}
\label{sec:eval_scheme}

In order to measure the puff detection performance (EX-I), we apply
the strict evaluation scheme presented in \cite{kyritsis2020data}. An
example of the evaluation scheme is presented in Figure
\ref{fig:detection1}. Essentially: a) only the first detected puff
within the duration of a GT interval is considered as a true positive
(TP), all subsequent ones count as false positives (FP), b) GT
intervals without detections count as false negative (FN) and c)
predictions outside GT intervals are considered as FP. It should be
noted that the evaluation scheme of \cite{kyritsis2020data} cannot
calculate true negatives (TN). However, at a window level we can
effectively measure TP/FP/FN and TN; i.e., by comparing the label
$y_{i}$ of each extracted window $\bm{W}_{i}$ with the GT. As a
result, we can calculate the weighted accuracy metric, defined as
$\frac{TP \cdot w + TN}{(TP + FN) \cdot w + FP + TN}$, using a weight
$w$ equal to $7.27$ (total time spend during smoking sessions divided
by the total time spend during puffs).

Regarding EX-II, a detected smoking session is considered a TP if it's
middle timestamp (calculated as $\frac{t_{i}^{s}+t_{i}^{e}}{2}$) is
within the duration of a GT interval; in any other case is considered
a FP. In addition, GT intervals without detections are considered as
FN. Figure \ref{fig:density} illustrates the aforementioned evaluation
scheme. Similar to EX-I, we also calculated the weighted accuracy for
EX-II using a weight equal to $13.76$. Finally, we calculated the
Jaccard Index (JI), defined as $\frac{|A \cap B|}{|A \cup B|}$, where
$A$ and $B$ are the intervals of the true and the predicted smoking
sessions, respectively.

\begin{table}
\caption{\small{EX-I/-II results.}}
\label{table:results}
\resizebox{0.50\textwidth}{!}{\begin{tabular}{l|l| c c c c c}
\toprule
\textbf{Experiment} & \textbf{Algorithm} \ & \textbf{W. Acc} & \textbf{Prec} & \textbf{Rec} & \textbf{F1-score} & \textbf{JI} \\
\midrule
\multirow{4}{*}{EX-I}&{\cite{shoaib2016hierarchical} with RF } & 0.894  & 0.834 & 0.840 & 0.837 & N/A \\
& {\cite{shoaib2016hierarchical} with SVM } & 0.876 & 0.836 & 0.815 & 0.825 & N/A \\
& {\cite{shoaib2016hierarchical} with DT } & 0.730  & 0.478 & 0.960 & 0.638 & N/A \\
& {Proposed} & \textbf{0.915} & 0.921 & 0.811 & \textbf{0.863} & N/A \\
\midrule
EX-II & {Proposed} & 0.968 & 0.837 & 0.923 & 0.878 & 0.604 \\
\bottomrule
\end{tabular}}
\end{table}



\section{Results}
\label{sec:res}

The obtained results showcase the high potential of our approach; both
towards the detection of individual puffs (upper part of Table
\ref{table:results}), as well as for the temporal localization of
smoking events in-the-wild (lower part of Table
\ref{table:results}). More specifically, regarding EX-I, the proposed
approach achieves a weighted accuracy of $0.915$ and an F$1$-score of
$0.863$ using the stricter evaluation scheme of
\cite{kyritsis2020data} (against $0.894$ and $0.837$ obtained by
\cite{shoaib2016hierarchical}). Concerning EX-II, our approach achieves
an F$1$-score/weighted accuracy/JI equal to $0.878$/$0.968$/$0.604$
which indicates that smoking sessions can be effectively detected
under in-the-wild conditions.


\section{Conclusions}
\label{sec:conclusions}

In this paper we present a two-step, bottom-up method towards the
in-the-wild monitoring of smoking behavior. LOSO experimental results
using our realistic SED and SED-FL datasets reveal the high potential
of our approach towards the detection of puffs and the localization of
smoking sessions during the day, under in-the-wild conditions.



\section{Acknowledgments}
\label{sec:ack}
The work leading to these results has received funding from the EU
Commission under Grant Agreement No. 965231, the REBECCA H2020 project
(\url{https://rebeccaproject.eu/}).

\bibliographystyle{IEEEtran} \bibliography{root}
\end{document}